\documentclass[reprint,
preprintnumbers,
nofootinbib,
nobibnotes,
superscriptaddress,
 amsmath,amssymb,
 aps,
prd,
]{revtex4-2}

\usepackage{graphicx, braket, orcidlink, algorithm2e, amssymb, amsmath, dsfont, adjustbox, multirow, lineno, nccmath, scalerel, xcolor, mathtools, booktabs}
\PassOptionsToPackage{breaklinks=true,colorlinks=true}{hyperref}
\PassOptionsToPackage{hyphens}{url}

\usepackage[normalem]{ulem}

\usepackage{dcolumn}
\usepackage{bm}
\usepackage{tikz}
\usetikzlibrary{quantikz2}

\usepackage{physics} 

\hypersetup{
    unicode=true,
    pdftoolbar=true,
    pdfmenubar=true,
    pdffitwindow=false,
    pdfstartview={fitH},
    pdftitle={TIMES},
    pdfauthor={B.~Sambasivam, K.~Sherbert, K.~Shirali, N.J.~Mayhall, E.~Barnes, S.E.~Economou},
    pdfsubject={TEPID},
    pdfcreator={B.~Sambasivam, K.~Sherbert, K.~Shirali, N.J.~Mayhall, E.~Barnes, S.E.~Economou},
    pdfproducer={},
    pdfkeywords={},
    pdfnewwindow=true,
    colorlinks=true,
    breaklinks=true,
    linkcolor=blue,
    citecolor=magenta,
    filecolor=magenta,
    urlcolor=blue
}

\begin{document}


\title{TIMES-ADAPT: A Quantum algorithm for real-time evolution in low-energy subspaces using fixed-depth circuits}

\author{Bharath Sambasivam\orcidlink{0000-0002-5765-9469}}
\email{sbharath@vt.edu}
\affiliation{Department of Physics, Virginia Tech, Blacksburg, VA 24061, USA}
\affiliation{Virginia Tech Center for Quantum Information Science and Engineering, Blacksburg, VA 24061, USA}

\author{Kyle Sherbert\orcidlink{0000-0002-5258-6539}}
\affiliation{Department of Physics, Virginia Tech, Blacksburg, VA 24061, USA}
\affiliation{Department of Chemistry, Virginia Tech, Blacksburg, VA 24061, USA}
\affiliation{Virginia Tech Center for Quantum Information Science and Engineering, Blacksburg, VA 24061, USA}

\author{Karunya Shirali\orcidlink{0000-0002-2006-2343}}
\affiliation{Department of Physics, Virginia Tech, Blacksburg, VA 24061, USA}
\affiliation{Virginia Tech Center for Quantum Information Science and Engineering, Blacksburg, VA 24061, USA}

\author{Nicholas J. Mayhall\orcidlink{0000-0002-1312-9781}}
\affiliation{Department of Chemistry, Virginia Tech, Blacksburg, VA 24061, USA}
\affiliation{Virginia Tech Center for Quantum Information Science and Engineering, Blacksburg, VA 24061, USA}

\author{Edwin Barnes\orcidlink{0000-0002-0982-9339}}
\affiliation{Department of Physics, Virginia Tech, Blacksburg, VA 24061, USA}
\affiliation{Virginia Tech Center for Quantum Information Science and Engineering, Blacksburg, VA 24061, USA}

\author{Sophia E. Economou\orcidlink{0000-0002-1939-5589}}
\email{economou@vt.edu}
\affiliation{Department of Physics, Virginia Tech, Blacksburg, VA 24061, USA}
\affiliation{Virginia Tech Center for Quantum Information Science and Engineering, Blacksburg, VA 24061, USA}

\date{\today}

\begin{abstract}
We propose a new variational quantum algorithm, which we refer to as TIMES-ADAPT, that prepares time-evolved states in a low-energy or symmetric subspace of a time-independent Hamiltonian on a quantum computer. Using a specially trained unitary that diagonalizes the Hamiltonian in a subspace, we construct fixed-depth circuits for real-time evolution in the subspace, where time only enters as a circuit parameter. We present two versions of the algorithm depending on whether the initial state is specified in the energy eigenbasis or computational basis. We consider two important applications of our methods: wave packet evolution and energy transport in spin systems. We benchmark our algorithms using variants of the Heisenberg XXZ model.
\end{abstract}

\maketitle


\section{Introduction}\label{sec:Intro}

Simulating the real-time dynamics of Hamiltonians has a wide range of important applications, including studying reactions in chemistry~\cite{KassalNASProceedings2008,ChanScience2023,KaleJPhysChemLett2024}, scattering in spin systems~\cite{GustafsonPRD2019,Yeter-AydenizNJoP2021} and high energy physics~\cite{JordanQuantum2018,BanulsEPJD2020,DavoudiQuantum2024,Bennewitz2024,Chai2025}, thermalization of many body systems~\cite{YangPRXQuantum2023,HahnPRX2024,PerrinNatrureComm2025}, dynamics of Floquet systems~\cite{EcksteinNPJQI2024,SekiPRR2025}, probing quantum chaos~\cite{JunPRX2017,AnandScientificReports2024,AsaduzzamanPRD2024}, and studying thermal or energy transport in materials~\cite{KarraschPRB2012,BulchandaniPNAS2023}. Using classical techniques for this is generally intractable, due to the exponential cost of storing and keeping track of the evolution of an entangled many-body quantum state. Tensor networks have emerged as a valuable classical technique to study real-time evolution (see Ref.~\cite{OrusNatureReviews2019} for a review), but are only effective for short times, when the entanglement in the system is not too large. Quantum computers offer a natural pathway around this problem, most straightforwardly for static Hamiltonians~\cite{FeynmanIJoTP1982}.

Existing quantum methods for real-time Hamiltonian simulation include product formulae~\cite{SuzukiJMP1991, BerryCMP2007, WiebeJPhysA2011, PoulinPRL2011, ChildsPhysRevX2021,CampbellPRL2019,ChildsQuantum2019}, including improved error bounds for low-energy initial states~\cite{SahinogluNPJQI2021}, linear combination of unitaries~\cite{Childs2012, ChakrabortyQuantum2024}, quantum random walks~\cite{ChildsPRL2009,ChildsCMP2009,BerryQIC2012}, qubitization~\cite{LowQuantum2019}, decompositions of dynamical Lie algebras~\cite{KokcuPRL2022, WiersemaNPJQI2024}, Krylov subspace methods~\cite{StairJCTC2020,EpperlySIAMJ2022,CortesPRA2022}, and variational methods~\cite{YuanQuantum2019,CirstoiuNPJQI2020,YaoPRXQ2021,BenedettiPRR2021,LinPRXQ2021,GibbsNPJQI2022,BarisonPRR2022,LinteauPRR2024,Zhang2025}. Lower-order product formulae and variational methods are the most amenable to implementation in the near term, given their low resource costs. However, product formulae at low truncation orders entail errors that become large at late times. Trotterization also breaks Hamiltonian symmetries, and as a result, ends up introducing errors stemming from parts of Hilbert space that do not belong in the same symmetry subspace as the initial state. Existing variational methods that use cost functions inspired by Mclachlan’s variational principle~\cite{YuanQuantum2019,BarisonQuantum2021} also require a discretization of time for numerics, thereby introducing errors that scale with time. Moreover, variational methods that use static ans\"atze also tend to yield deeper circuits and suffer from optimization issues. This can in part be alleviated by using adaptive constructions~\cite{YaoPRXQ2021,LinteauPRR2024,Zhang2025}. In Refs.~\cite{CirstoiuNPJQI2020,GibbsNPJQI2022}, a trotterized unitary evolution of the input state for a short period of time is considered and variationally diagonalized to target eigenstates of the Hamiltonian that have overlap with the initial state. The diagonalizing unitary is then used to fast-forward the initial state. This effectively yields a fixed-depth circuit, where time enters as a circuit parameter. This circuit is capable of maintaining good fidelity with exact evolution for longer times compared to what is achievable with product formulae. Nevertheless, the usage of a trotterized approximation in the beginning introduces errors that eventually become significant in the large-time limit. In Ref.~\cite{CortesPRA2022}, Krylov subspace methods with multiple reference states are used to fast-forward real-time dynamics of an initial state. In spite of this, the error does scale with time, requiring a larger number of reference states to maintain good fidelity at late times. Approaches that involve decompositions of the Hamiltonian’s dynamical Lie algebra~\cite{KokcuPRL2022} can yield truly fixed-depth circuits. However, the resource requirements are expected to scale unfavorably for systems with interacting fermions~\cite{WiersemaNPJQI2024,Aguilar2024}.

In this work, we present a new quantum algorithm that we call TIMES-ADAPT that prepares states that are time-evolved with respect to time-independent Hamiltonians up to \emph{arbitrarily large times} using fixed-depth circuits. We begin by training a parametrized circuit to diagonalize the Hamiltonian in the subspace of interest. We achieve this via a modular ansatz that prepares a thermal Gibbs state of the Hamiltonian using TEPID-ADAPT~\cite{Sambasivam2025}, where the free energy is the cost function. TEPID-ADAPT produces both the low-energy eigenvalues as well as a basis-change circuit that maps computational basis states to low-energy eigenstates---outputs that are key ingredients for TIMES-ADAPT. In this work, we also extend this to symmetry subspaces by engineering the pool of operators used in the ADAPT-VQE subroutine to stay within the symmetry subspace of interest. We present two versions of TIMES-ADAPT that differ according to whether the initial state is specified in either the energy eigenbasis or the computational basis. In both cases, TIMES-ADAPT yields a fixed-depth circuit, where time enters as a parameter in the state preparation step. If the initial state is entirely in the subspace found using TEPID-ADAPT, the only source of errors is the variational preparation of the basis-change unitary. Because it avoids trotterization completely, TIMES-ADAPT does not suffer from an accumulation of errors at large times, and can entirely avoid breaking system symmetries. It also suffers no obvious restriction from the scaling of the size of the dynamical Lie algebra of the system of interest.

We demonstrate TIMES-ADAPT on two applications: wave-packet evolution and energy transport in spin systems. Wave packet evolution is crucial for studying scattering in models on a lattice such as spin systems and lattice gauge theories. Using our methods, we can study the evolution of wave packets in symmetry subspaces with fixed-depth circuits. Studying thermal and energy transport in materials is an interesting problem in non-equilibrium statistical physics that has implications for thermalization~\cite{KarraschPRB2012,BulchandaniPNAS2023}. The late-time dynamical behavior of thermal perturbations can help categorize systems into universality classes. The perturbation and its evolution live primarily in the low-energy sector, making it a great application of our fixed-depth algorithms.

The paper is organized as follows. In Sec.~\ref{sec:TEPID-Review}, we review TEPID-ADAPT~\cite{Sambasivam2025} as a means of diagonalizing the Hamiltonian in a low-energy or symmetry subspace. In Sec.~\ref{sec:OurMethod}, we introduce our algorithm TIMES-ADAPT to study real-time dynamics of time-independent Hamiltonians in a subspace using fixed-depth circuits. In Sec.~\ref{sec:Results}, we apply TIMES-ADAPT to the ferromagnetic phase of the Heisenberg XXZ model and compare our results with exact diagonalization. In Sec.~\ref{sec:Conclusions}, we conclude and offer future directions to pursue. In Appendix~\ref{app:Uses}, we present two applications of TIMES-ADAPT: wave-packet evolution and energy transport. Appendices~\ref{app:t0prep} and~\ref{app:ErrorEstimates} contain further details of the algorithms.

\section{Review of TEPID-ADAPT}\label{sec:TEPID-Review}
In this section, we briefly review TEPID-ADAPT~\cite{Sambasivam2025}, the method we use to diagonalize the low-energy subspace of interest via preparing a low-temperature Gibbs state
\begin{equation}
    \rho_G\equiv \frac{e^{-\beta\,{H}}}{Z},\qquad Z=\Tr(e^{-\beta\,{H}}),
\end{equation}
where $\beta$ is the inverse temperature. 
\begin{figure}[ht!]
    \centering
    \begin{quantikz}[classical gap=0.07cm]
        \lstick{$\ket{0}^{\otimes N_s}$} & \qwbundle{N_s} & \gate[2]{U_m(\vec{\mu})}&\slice{prep} & & \gate{V_A(\vec{\theta})}&\rstick{$\approx\rho_G$}\\
        \lstick{$\ket{0}^{\otimes N_a}$} & \qwbundle{N_a} & & & & & \trash{\text{trace}}
    \end{quantikz}
    \caption{General block diagram for the variational ansatz of TEPID-ADAPT. $U_m(\vec{\mu})$ prepares $\rho_m$ on the system register (top), as indicated by the red line. $V_A(\vec{\theta})$ is an adaptively generated unitary on the system register that approximately evolves $\rho_m$ to the target Gibbs state.}
    \label{fig:BlockAnsatz}
\end{figure}
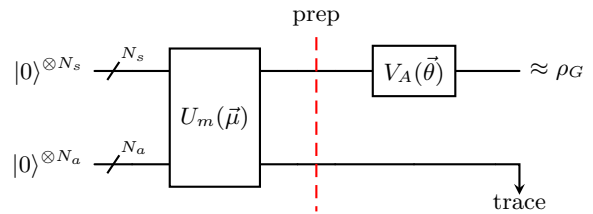
TEPID-ADAPT uses the blocked parametrized ansatz shown in Fig.~\ref{fig:BlockAnsatz} to target $m$ low-lying eigenstates of the Hamiltonian simultaneously. The static part of the ansatz $U_m(\vec{\mu})$ uses an ancillary register of size $N_a\geq\lceil\log_2m\rceil$ to prepare a parametrized diagonal density matrix with support on $m$ computational basis elements on the system register:
\begin{equation}
    \rho_m=\sum_{k=1}^m\mu_k\ket{c_k}\bra{c_k},
\end{equation}
where $\left\{\ket{c_k}\right\}_{k=1}^m$ are the chosen computational basis states. Upon convergence, the adaptively generated unitary $V_A$ maps this computational basis subspace to a subspace $\mathbf{S}$ spanned by $m$ low-energy eigenstates $\{\ket{\psi_k}\}_{k\in\mathbf{S}}$:
\begin{equation}\label{eq:Vperpdef}
    V_A\equiv\sum_{k\in\mathbf{S}}\ket{\psi_k}\bra{c_k}+V_{\perp},
\end{equation}
where $V_{\perp}$ is an operator that annihilates the computational basis states $\left\{\ket{c_k}\right\}_{k=1}^m$. Note that the action of $V_A$ on $\left\{\ket{c_k}\right\}_{k=m+1}^{2^N_s}$ is not specified in Eq.~\eqref{eq:Vperpdef}. In Ref.~\cite{Sambasivam2025}, the target subspace was spanned by the absolute lowest $m$ eigenstates of $H$ because the goal was to prepare the low-temperature Gibbs state. In this work, we consider more general low-energy subspaces, $\mathbf{S}$. For example, one could target the eigenstates that belong to a symmetry subspace using a judicious choice of computational basis states and pool operators used to adaptively build $V_A$. In this case, the state prepared using TEPID-ADAPT would be a low-temperature Gibbs state of the Hamiltonian projected onto the symmetry subspace:
\begin{equation}
    \rho_G\equiv \frac{e^{-\beta\,{H_{\mathbf{S}}}}}{Z_{\mathbf{S}}},\qquad Z_{\mathbf{S}}=\Tr(e^{-\beta\,{H_{\mathbf{S}}}}),
\end{equation}
where $H_{\mathbf{S}}=\mathbb{P}_{\mathbf{S}}\,H\,\mathbb{P}_{\mathbf{S}}$, with $\mathbb{P}_{\mathbf{S}}$ being the projector onto the symmetry subspace $\mathbf{S}$. We consider such an example in Appendix.~\ref{app:Uses}, where we target eigenstates in the single-magnon subspace of a spin system. 

Note that in this work, we are preparing the Gibbs state using TEPID-ADAPT solely as a means to train a circuit that diagonalizes the Hamiltonian in the desired low-energy subspace. One could alternatively use an approach like concurrent VQE~\cite{XuPRA2023} to target multiple low-lying eigenstates simultaneously; however, in this approach measurements on each eigenstate are required to obtain the energy differences between the eigenstates. In contrast, with TEPID-ADAPT we can forgo this overhead and directly obtain them from the converged parameters as a result of the particular parametrization and optimization used to prepare the Gibbs state:
\begin{equation}\label{eq:EnergyDiffs}
    \Delta E_k = \frac{1}{\beta}\log\frac{\mu_1}{\mu_k}\,\quad\forall\,\,k\in\mathbf{S},
\end{equation}
where $\Delta E_k\equiv E_k-E_1$. These energy differences set the relative phases between the various eigenstates during real-time evolution. Once we have trained the adaptively generated unitary $V_A$, we can study the real-time dynamics of any arbitrary state in the prepared low-energy subspace using fixed-depth circuits, as we show in Sec.~\ref{sec:OurMethod}. 


\section{TIMES-ADAPT}\label{sec:OurMethod}
\begin{figure*}[t!]
    \includegraphics[width=\linewidth]{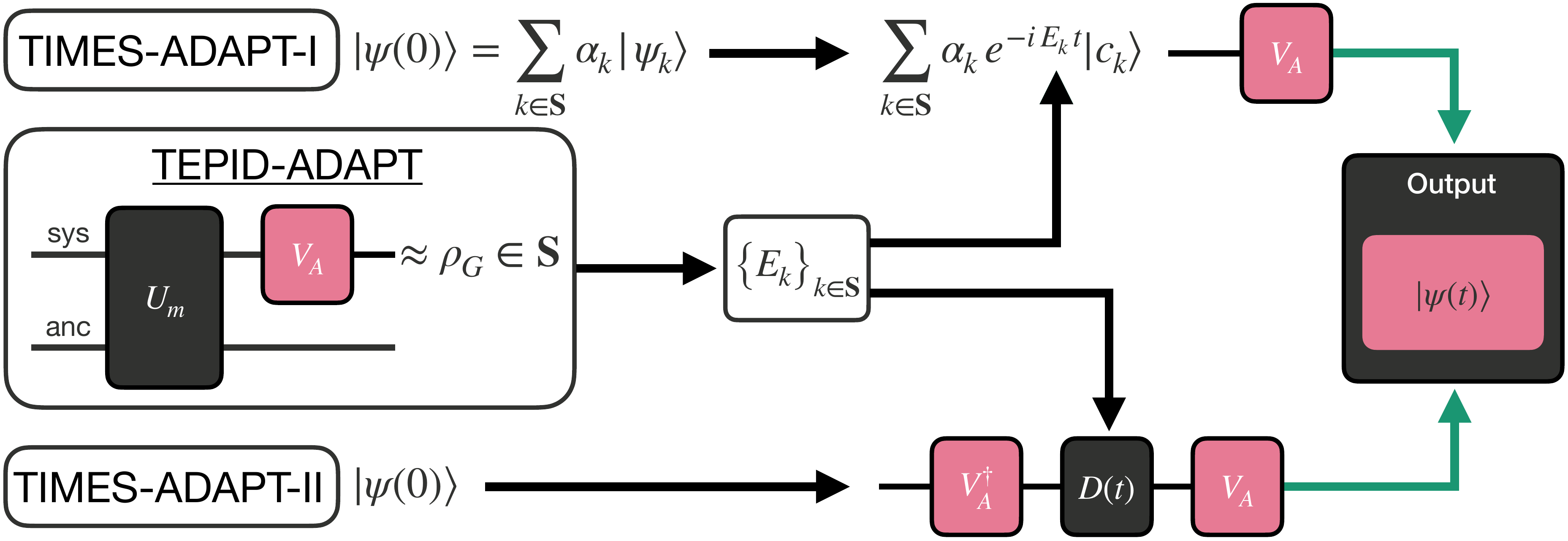}
    \caption{A diagrammatic workflow for TIMES-ADAPT. We first use TEPID-ADAPT to train an adaptively generated circuit $V_A$ to diagonalize the Hamiltonian in a low-energy subspace. We also readily obtain the energy differences from this procedure. We propose two related versions of our algorithm to perform time-evolution of a state in the low-energy subspace with a fixed-depth circuit. In TIMES-ADAPT-I, we create a particular linear combination of computational basis states, and use $V_A$ to obtain the time-evolved state $\ket{\psi(t)}$. In TIMES-ADAPT-II, we compile an effective unitary that is equivalent to the evolution operator in the low-energy subspace, and then apply it to the initial state to obtain the time-evolved state $\ket{\psi(t)}$.}
    \label{fig:SummaryChart}
\end{figure*}

In this section, we introduce two related versions of our quantum algorithm for simulating the real-time dynamics of many-body quantum systems governed by time-independent Hamiltonians on quantum computers. We collectively the algorithm TIMES-ADAPT, which stands for \textbf{T}hermal \textbf{I}nspired \textbf{M}ethods for \textbf{E}volution in a \textbf{S}ubspace. The input we assume for these algorithms is a circuit that prepares the initial state we would like to evolve, $\ket{\psi(0)}$. We will denote the subspace found by TEPID-ADAPT by $\mathbf{S}$. We provide an analysis of the errors incurred if the initial state is not entirely in $\mathbf{S}$ in Appendix~\ref{app:ErrorEstimates}. Going forward, we will only work with the system register---we can discard the ancillary register used for preparing the thermal state. In Fig.~\ref{fig:SummaryChart}, we show a diagrammatic workflow of the two versions of TIMES-ADAPT.

\subsection{TIMES-ADAPT-I}

In the first version of TIMES-ADAPT, which we call TIMES-ADAPT-I, we assume that the initial state is specified in terms of the truncated eigenbasis $\{\ket{\psi_k}\}_{k\in\mathbf{S}}$:
\begin{equation}\label{eq:t0StateExpansion}
	\ket{\psi(0)}=\sum_{k\in\mathbf{S}}\alpha_k\,\ket{\psi_k}.
\end{equation}
In the case where the coefficients $\alpha_k$ are not known \textit{a priori}, we propose a method to find them in Appendix~\ref{app:FindCoeffs}. 

First, we prepare the superposition state
\begin{equation}\label{eq:PrepState}
	\ket{\chi(t)}=\sum_{k\in\mathbf{S}}\alpha_k\,e^{-i\,E_k\,t}\ket{c_k},
\end{equation}
where $\{E_k\}_{k\in\mathbf{S}}$ are the energies corresponding to the eigenstates of $H$ in $\mathbf{S}$, $\{\ket{c_k}\}_{k\in\mathbf{S}}$ are the computational basis elements we used to prepare the Gibbs state, and $t$ is the time we aim to evolve to. Note that an overall phase in Eq.~\eqref{eq:PrepState} can be factored out, so that the time-evolved state only depends on the energy differences in Eq.~\eqref{eq:EnergyDiffs}. The specific procedure for this state preparation is provided in Appendix~\ref{app:t0prep}.
For the purposes of this state preparation, we assume that $\sum_{k\in\mathbf{S}}\lvert\alpha_k\rvert^2=1$, i.e., that the initial state $\ket{\psi(0)}$ lies completely in the subspace $\mathbf{S}$.
When this is not the case, the $\alpha_k$ are renormalized to ensure $\ket{\chi(t)}$ is a normalized state.
We discuss the errors incurred by this normalization at the end of this section.

Next, we act the unitary $V_A$ on the state $\ket{\chi(t)}$ to obtain
\begin{equation}
	\ket{\psi(t)}\equiv V_A\ket{\chi(t)}=\sum_{k\in\mathbf{S}}\alpha_k\,e^{-i\,E_k\,t}\ket{\psi_k}.
\end{equation}
This is precisely the state one would get by evolving the initial state $\ket{\psi(0)}$ by an \textit{arbitrary} amount of time $t$ under the Hamiltonian $H$. We have used a \textit{fixed depth} circuit that involves the preparation of $\ket{\chi(t)}$ in Eq.~\eqref{eq:PrepState}, followed by the action of $V_A$.

In case the initial state $\ket{\psi(0)}$ is not entirely in the subspace found by TEPID-ADAPT, the fidelity of the evolved state with exact diagonalization is a constant (time-independent) factor less than unity. This is due to the fact that we are effectively evolving the projection of $\ket{\psi(0)}$ onto the subspace found by TEPID-ADAPT. The norm of the state after the projection does not change with time, leading to the constant reduction in fidelity. We show this in detail in Appendix~\ref{app:ErrorEstimates}. We demonstrate TIMES-ADAPT-I in Sec.~\ref{sec:Results} for the Heisenberg XXZ model in the antiferromagnetic phase. We also present the evolution of a wave packet as an application of TIMES-ADAPT-I in Appendix~\ref{app:Uses}.


\subsection{TIMES-ADAPT-II}

The second version of the algorithm, which we call TIMES-ADAPT-II, directly evolves an initial state specified in the computational basis under the Hamiltonian truncated to the relevant low-energy subspace. The effective Hamiltonian $\tilde{H}$ of rank $m$ has the energy spectrum
\begin{equation}
    \vec{\Lambda}=\left(E_1,\cdots,E_m,0,\cdots,0\right),
\end{equation}
where $m$ energies are identical to the original Hamiltonian, $H$, corresponding to the eigenstates that live in the low-energy subspace of interest. The rest of the energies are effectively zero, as a result of the truncation scheme used in TEPID-ADAPT. The action of $\tilde{U}(t)=\exp\left(-i\,\tilde{H}\,t\right)$ on a vector in the truncated eigenspace is identical to that of the full evolution unitary $\exp\left(-i\,H\,t\right)$.

To compile $\tilde{U}(t)$, we first construct a general diagonal unitary
\begin{equation}\label{eq:DtTIMES-II}
    D(t)\equiv\sum_{k=1}^{2^{N_s}}\exp(-i\,\Lambda_k\,t)\ketbra{c_k}{c_k}
\end{equation}
from a sequence of $m$ multi-controlled phase gates applied to the $N_s$ system qubits.
Next, we conjugate with the converged basis-change unitary $V_A$ that we obtained from TEPID-ADAPT to get the required effective evolution unitary:
\begin{align}\label{eq:TIMES-ADAPT-II-Unitary}
\begin{split}
    \tilde{U}(t)&=V_A\,D(t)\,V_A^{\dagger} \\
    &= \sum_{k=1}^{m}e^{-i\,E_k\,t}\ketbra{\psi_k}{\psi_k}+\sum_{k=m+1}^{2^{N_s}}\ketbra{w_k}{w_k},
\end{split}
\end{align}
where $\ket{w_k}\equiv V_{\perp}\ket{c_k}$, and $V_{\perp}$ was introduced in Eq.~\eqref{eq:Vperpdef}. The unitary $\tilde{U}(t)$ can be applied to any initial state, while being agnostic to its coefficients in the truncated eigenspace. Note that the results will be accurate only when the initial state lies entirely in the low-energy subspace $\mathbf{S}$. In case the initial state $\ket{\psi(0)}$ is not entirely in $\mathbf{S}$, the fidelity of the evolved state oscillates with frequencies set by the energies of the eigenstates that are not in $\mathbf{S}$. We show this in Appendix~\ref{app:ErrorEstimates}. We demonstrate TIMES-ADAPT-II in Sec.~\ref{sec:Results} for the Heisenberg XXZ model in the antiferromagnetic phase. We also present the dynamics of energy transport as an application of TIMES-ADAPT-II in Appendix~\ref{app:Uses}.


\section{Numerical Results}\label{sec:Results}
In this section, we showcase both versions of TIMES-ADAPT using the anti-ferromagnetic phase ($J_z>1.0$) of the Heisenberg XXZ model with six qubits, described by the Hamiltonian:

\begin{multline}\label{eq:XXZHam}
    H_{\text{XXZ}}(J_z)=\sum_{k=1}^{5}\left(\sigma_{k}^x\sigma_{k+1}^x+\sigma_{k}^y\sigma_{k+1}^y+J_z\sigma_{k}^z\sigma_{k+1}^z\right),
\end{multline}
where, $J_z$ is the coupling in the $Z$-direction, and $\sigma^{x,y,z}_j$ are the Pauli operators acting on qubit $j$.
\begin{figure*}[t!]
    \includegraphics[width=\linewidth]{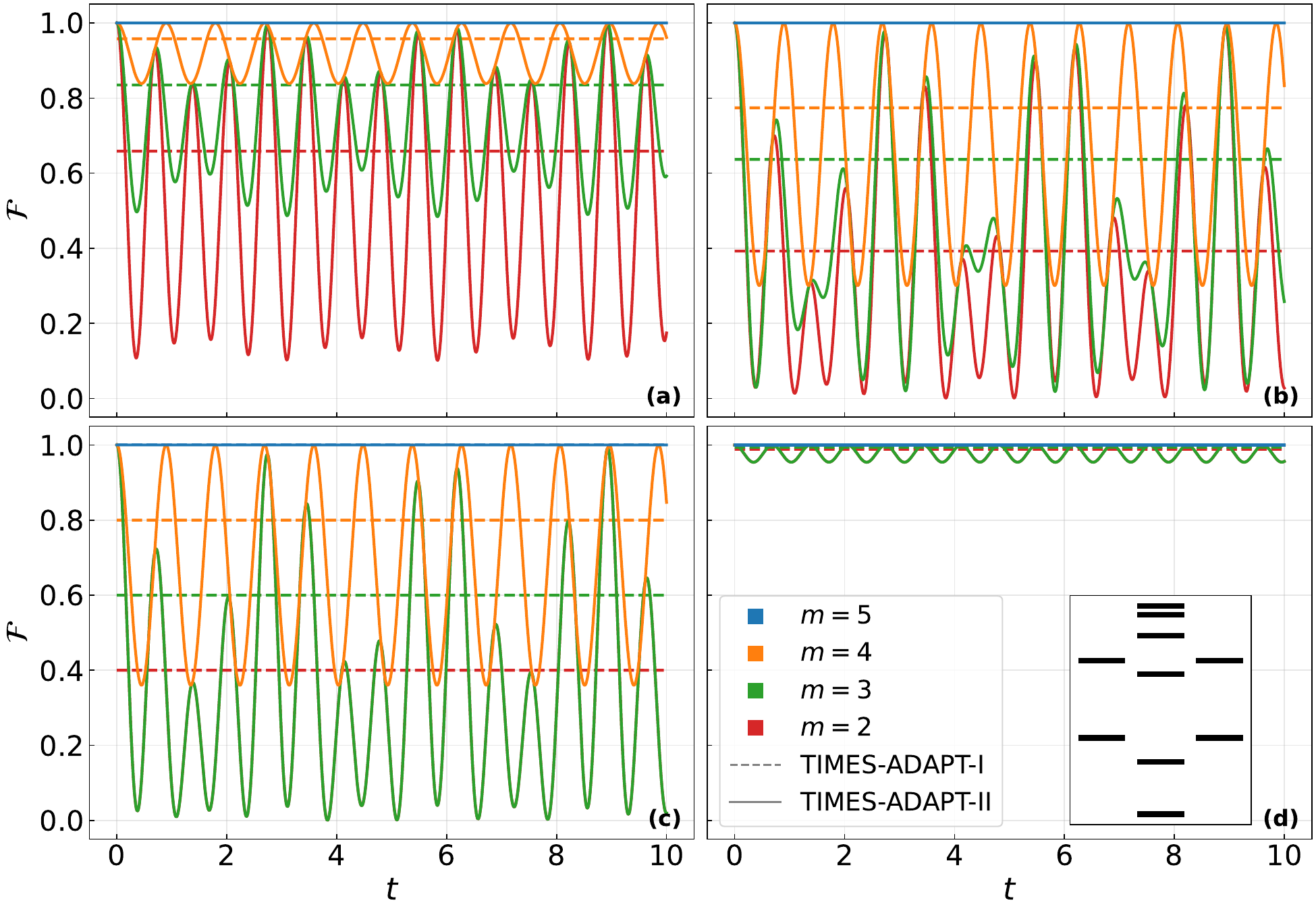}
    \caption{The fidelity of various time-evolved states using TIMES-ADAPT-I (dashed) and TIMES-ADAPT-II (solid) with exact diagonalization. The initial state in \textbf{(a)}, \textbf{(b)}, \textbf{(c)} are assumed to be in the span of the first five eigenstates of $H_{\text{XXZ}}$. In \textbf{(d)}, the initial state is a linear combination of all the eigenstates with their Boltzmann weights as coefficients. The various colors corresponds to which portion of the subspace is known via TEPID-ADAPT, as indicated. Panels \textbf{(a)}, \textbf{(b)} show results for random initial states, whereas \textbf{(c)} and \textbf{(d)} show results for a uniform and exponential distribution of coefficients, respectively.} 
    \label{fig:M-I-II-All}
\end{figure*}

We use TEPID-ADAPT to adaptively train a circuit to find a unitary $V_A$ that diagonalizes the Hamiltonian in the subspace spanned by the lowest five eigenstates of $H_{\text{XXZ}}(1.5)$. We do this via preparing the Gibbs state with $\beta=2.0$\footnote{Since the quality of the rank-truncated Gibbs state does not matter for the purpose of finding a fixed number of low-energy eigenstates using TEPID-ADAPT, the temperature is a hyperparameter, and a choice of any moderate temperature is appropriate.}. We then consider the real-time evolution of various states in the same low-energy subspace. To test the performance of TIMES-ADAPT when the initial state is not entirely in the subspace, we consider cases where the basis-change unitary $V_A$ diagonalizes $H_{\text{XXZ}}(1.5)$ in lower-dimensional subspaces. We consider various initial states defined by their coefficients in the truncated eigenbasis:
\begin{equation}
    \ket{\psi(0)}=\frac{1}{\mathcal{N}}\sum_{k=1}^5\alpha_k\ket{\psi_k}.
\end{equation}

In Fig.~\ref{fig:M-I-II-All}, we plot the fidelity of the evolved state (computed relative to the result from exact diagonalization) obtained using different numbers $m$ of the low-energy eigenstates retained in the the TEPID-ADAPT subroutine. As indicated in the plot legends, the dashed and solid lines in Fig.~\ref{fig:M-I-II-All} correspond to real-time evolution using TIMES-ADAPT-I and TIMES-ADAPT-II, respectively. The inset shows an illustration of what the low-energy eigenspectrum of $H_{\text{XXZ}}(1.5)$ looks like. We consider three types of initial states/coefficients:
\begin{itemize}
    \item In Figs.~\ref{fig:M-I-II-All} \textbf{(a)}, \textbf{(b)}, we consider two sets of random coefficients $\{\alpha_k\}_{k=1}^5$. Here, $\mathcal{N}=1$.
    \item In Fig.~\ref{fig:M-I-II-All} \textbf{(c)}, we consider a state that is in an equal superposition of the first five eigenstates (i.e., $\alpha_k=1/\sqrt{5}$). Here, $\mathcal{N}=1$.
    \item In Fig.~\ref{fig:M-I-II-All} \textbf{(d)}, we consider a case where the coefficients follow an exponential fall-off with the energy of the corresponding eigenstate (i.e., $\alpha_k=\exp(-E_k/2)/Z$, $Z=\sqrt{\sum_{k=1}^{2^{N_s}}\exp(-E_k)}$). Here, $\mathcal{N}<1$ because we consider non-zero coefficients corresponding to every eigenstate of $H$.
\end{itemize}

The loss in fidelity of the state evolved using TIMES-ADAPT-I when $m<5$ is constant, as mentioned in Sec.~\ref{sec:OurMethod}. The fidelity is determined by the norm of the initial state after it is projected onto the corresponding $m$-dimensional subspace (see Appendix~\ref{app:ErrorEstimates}). This is why the fidelities are equally spaced for the different values of $m$ when the initial state is an equal superposition of the eigenstates in the $m=5$ subspace (Fig.~\ref{fig:M-I-II-All}\textbf{(c)}). 

For TIMES-ADAPT-II, on the other hand, the fidelity when $m<5$ is oscillatory with frequencies set by the energies of the eigenstates outside the $m<5$ subspace. This can be seen in Fig.~\ref{fig:M-I-II-All}, where the fidelity corresponding to $m=4$ oscillates with a single frequency, corresponding to the energy difference between the fourth excited state and the ground state of $H_{\text{XXZ}}(1.5)$. For $m<4$, the oscillations of the fidelities are described by multiple frequencies, as expected.

TIMES-ADAPT-I is more appropriate when the initial state is specified in the eigenbasis of the Hamiltonian. A good application would be evolution of an initial state in a symmetry subspace. In Appendix~\ref{app:Uses}, we provide an explicit example of the evolution of a single-magnon wave packet in a system that preserves spin along the $Z-$ direction. On the other hand, TIMES-ADAPT-II is more appropriate when the initial state is specified in the computational basis, or if we only have black-box access to it. A good application would be the evolution of a low-energy initial state. In Appendix~\ref{app:Uses}, we showcase the example of energy transport in spin-systems, where the initial state is a local perturbation of the ground state of the model. The same can be extended to study thermal transport as well.

In terms of resources, TIMES-ADAPT-I uses an additional variational subroutine outlined in Appendix~\ref{app:FindCoeffs} to find the coefficients of the initial state in the truncated eigenbasis, but only needs one application of $V_A$ in the algorithm after the coefficients are known. In TIMES-ADAPT-II, there is no initial computational overhead, but it requires application of both $V_A$ and $V_A^{\dagger}$ in the algorithm. As a result, TIMES-ADAPT-I may be more amenable to near-term applications, since it yields shallower circuits, albeit at the cost of a subroutine to find the coefficients of the initial state in the truncated eigenbasis.


\section{Conclusions and Outlook}\label{sec:Conclusions}
In this article, we introduced TIMES-ADAPT, a variational algorithmic framework for real-time evolution under a Hamiltonian in a subspace using fixed-depth circuits. The algorithms use TEPID-ADAPT~\cite{Sambasivam2025} as a subroutine to train an adaptively generated circuit that simultaneously targets multiple eigenstates of a Hamiltonian in a subspace of interest, via preparing a Gibbs state. The converged parameters of this procedure also readily give us the energy gaps of the Hamiltonian in the subspace. The first version of our proposed algorithm, TIMES-ADAPT-I involves finding the expansion of the initial state in the truncated eigenbasis. Using these coefficients, dressed with the appropriate phases generated by the evolution of the individual eigenstates, we can prepare a superposition in the computational basis. Upon action by the basis-change unitary found using TEPID-ADAPT, we obtain the time-evolved state. In the second version, TIMES-ADAPT-II, we directly compile the effective time-evolution unitary in the subspace using the basis-change unitary. This allows us to evolve the initial state without knowledge of the coefficients in the truncated eigenbasis. This of course works best when the initial state is entirely in the subspace found by TEPID-ADAPT.

In Appendix~\ref{app:Uses}, we provide examples of where our methods could be applied. For TIMES-ADAPT-I, we consider the example of the evolution of a single-magnon Gaussian wave packet in the longitudinal-field Heisenberg XXZ model. The wave packet belongs to a symmetry subspace, where the total spin in the $Z$-direction is preserved. We can use TEPID-ADAPT to find the eigenstates that span this subspace, and use the basis-change unitary to find the time-evolved wave packet. This can be extended to study scattering of two wave packets. We will explore this in future work. For TIMES-ADAPT-II, we consider the example of energy transport in the staggered longitudinal-field Heisenberg XXZ model. We consider a localized perturbation on the ground state of the system and study its evolution using a fixed-depth circuit. The initial state is predominantly in the low-energy subspace, leading to a high fidelity that persists over arbitrarily long times.

A possible extension of this work would be studying real-time evolution in more complicated symmetry subspaces that does not fully coincide with a low-energy subspace. A good example of this is charge subspaces in high-energy physics models, where some configurations in a given charge subspace have higher energies. Similarly, in molecular problems, studying evolution in the subspace corresponding to a particular fixed particle number is of interest. These would involve careful choices of the initial computational subspace and pool operators for TEPID-ADAPT. We save this for future work, as well.


\section{Acknowledgments}\label{sec:Thanks}
We thank Norman Tubman, Tianci Zhou, Goksu Toga, and Henry Lamm for fruitful discussions. We thank Mafalda Ram\^{o}a and Priya Batra for useful comments during the development of TIMES-ADAPT. The Quantikz package~\cite{Kay2023} was used to generate the circuits in this paper. B. Sambasivam, K. Sherbert, K. Shirali, and S. E. Economou are supported by the U.S. Department of Energy, Office of Science, National Quantum Information Science Research Centers, Co-design Center for Quantum Advantage (C$^2$QA) under Contract No.DE-SC0012704.  N. J. Mayhall acknowledges funding from the Department of Energy (Award number: DE-SC0024619). E. Barnes acknowledges support from the Department of Energy (Award no. DE-SC0025430).


\bibliography{apssamp}


\appendix
\clearpage

\section{Applications}\label{app:Uses}
In this Appendix, we consider two applications of TIMES-ADAPT: wave-packet evolution and energy transport in spin systems. These applications serve to respectively illustrate the two versions of  TIMES-ADAPT introduced in Sec.~\ref{sec:OurMethod}.


\subsection{Wave-packet evolution}
It is important to find efficient ways to simulate the real-time dynamics of wave packets when studying scattering in spin systems. In this subsection, we will consider the evolution of a single-magnon Gaussian wave packet of a seven-qubit longitudinal-field Heisenberg XXZ model (LFXXZ) with open boundary conditions described by the Hamiltonian
\begin{multline}\label{eq:XXZHam1}
    H_{\text{LFXXZ}}(J_z, h_z)=\sum_{k=1}^{N_s-1}\left(\sigma_{k}^x\sigma_{k+1}^x+\sigma_{k}^y\sigma_{k+1}^y\right.\\ \left.+J_z\sigma_{k}^z\sigma_{k+1}^z\right)+h_z\sum_{k=1}^{N_s}\sigma_k^z,
\end{multline}
where $J_z$ is the coupling in the $Z$-direction, $h_z$ is the strength of the longitudinal magnetic field, $N_s$ is the number of spins in the chain, and $\sigma^{x,y,z}_j$ are the Pauli operators acting on qubit $j$. This Hamiltonian preserves the spin in the $Z$-direction:
\begin{equation}\label{eq:HTFXXZ}
    \left[S_z,H_{\text{LFXXZ}}(J_z,h_z)\right]=0,\quad S_z\equiv\sum_{j=1}^{N_s}\sigma^z_j.
\end{equation}
The single-magnon wave packet and its subsequent time evolution live in a symmetric subspace $\mathbf{S}_1$ with $\langle S_z\rangle=N_s-2$,
\begin{equation}
    \mathbf{S}_1\equiv\text{span}\left\{\sigma^x_1\ket{0}^{\otimes N_s},\cdots,\sigma^x_{N_s}\ket{0}^{\otimes N_s}\right\}.
\end{equation}
The single-magnon Gaussian wave packet localized at site $j_0$ with standard deviation $\lambda$ is
\begin{equation}
    \ket{\psi_{\text{G}}(j_0,\lambda,0)}=\frac{1}{\sqrt{\mathcal{N}}}\sum_{j=1}^{N_s}\exp\left(-\frac{\lambda^2}{2}(j-j_0)^2\right)\sigma^x_j\ket{0}^{\otimes N_s},
\end{equation}
where $\mathcal{N}\equiv\sum_{j=1}^{N_s}\exp(-\lambda^2(j-j_0)^2)$ is the normalization. To study the real-time evolution of this state, we first use TEPID-ADAPT to find the $N_s$ eigenstates $\left\{\ket{\psi_k}\right\}_{k\in\mathbf{S}_1}$ of $H_{\text{LFXXZ}}(J_z,h_z)$ that span $\mathbf{S}_1$. The appropriate choice of computational basis set for TEPID-ADAPT is $\left\{\ket{c_j}\right\}_{j=1}^{N_s}\equiv\left\{\sigma^x_j\ket{0}^{\otimes N_s}\right\}_{j=1}^{N_s}$. The appropriate operator pool to build the basis-change unitary $V_A$ in Fig.~\ref{fig:BlockAnsatz} consists of operators that preserve $S_z$. We use the qubit-excitation-based (QEB) pool~\cite{YordanovNatComm2021} here. The converged basis-change unitary maps the chosen computational basis set to the eigenstates of $H_{\text{LFXXZ}}$ that span $\mathbf{S}_1$:
\begin{equation}
    V_A(\vec{\theta}^*):\sigma_j^x\ket{0}^{\otimes N_s}\to\ket{\psi_j}\in\mathbf{S}_1,\quad\forall j=1,\cdots,N_s.
\end{equation}
Next, we find the coefficients $\{\alpha_k\}_{k\in\mathbf{S}_1}$ of $\ket{\psi_\text{G}(j_0,\lambda,0)}$ in the eigenbasis of the Hamiltonian,
\begin{equation}\label{eq:GWavePacket}
    \ket{\psi_{\text{G}}(j_0,\lambda,0)}=\sum_{k\in\mathbf{S}_1}\alpha_k\ket{\psi_k},
\end{equation}
using the method outlined in Appendix~\ref{app:FindCoeffs}. With the knowledge of $\{\alpha_k\}_{k\in\mathbf{S}_1}$, we can use TIMES-ADAPT-I to obtain the time-evolved wave packet at arbitrary times $t$ with a fixed-depth circuit.

\begin{figure*}
         \includegraphics[width=0.49\textwidth]{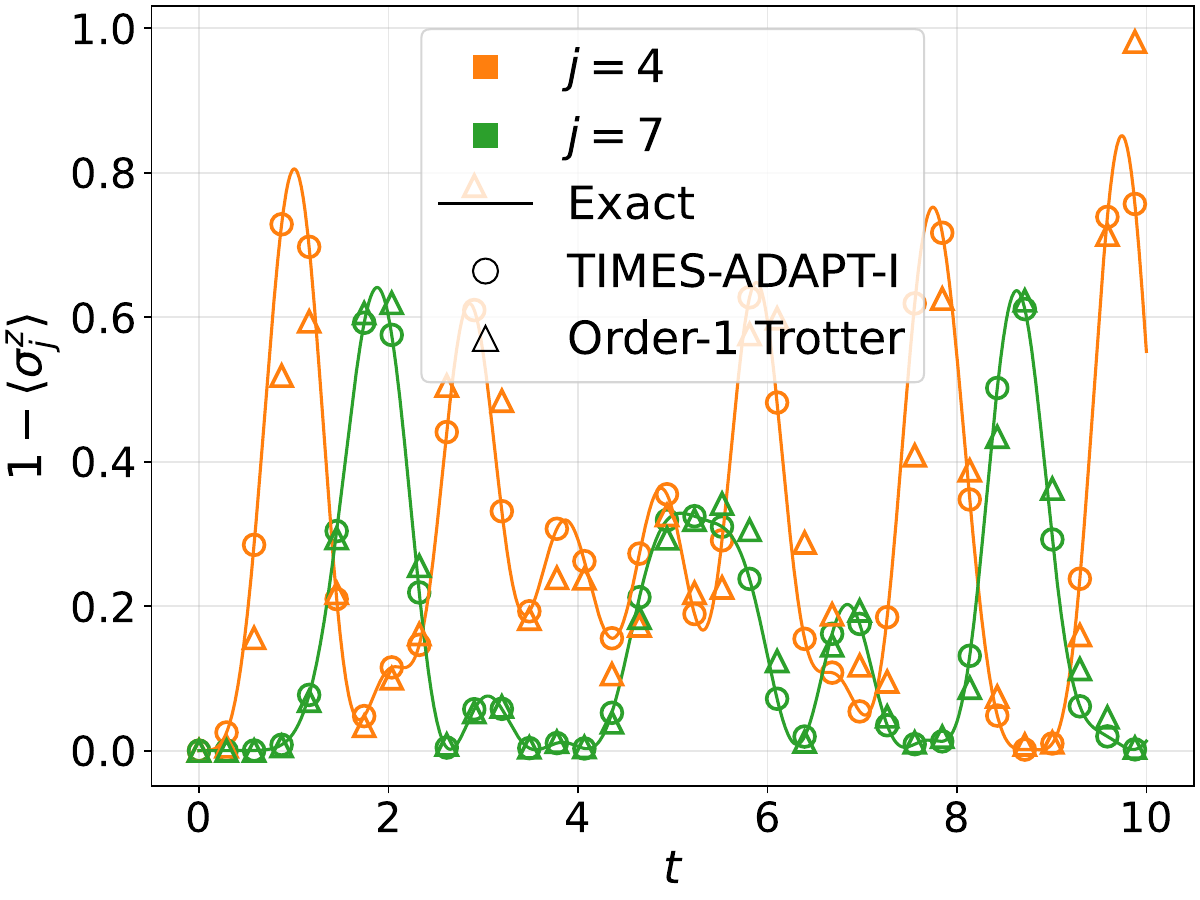}
         \label{fig:WavePacket_Obs}
     \hfill
         \includegraphics[width=0.49\textwidth]{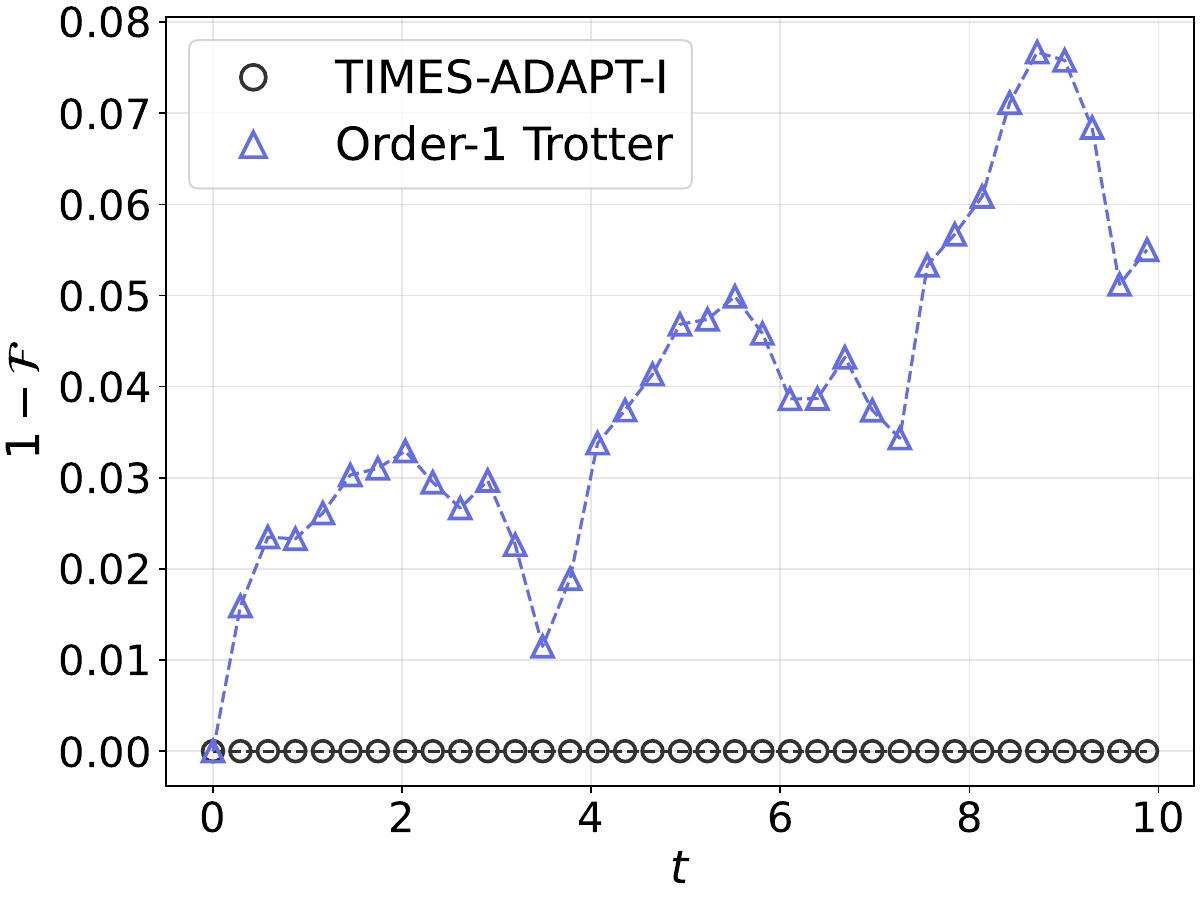}
         \label{fig:WavePacket_InF}
        \caption{Time evolution of a single-magnon Gaussian wave packet in the longitudinal-field XXZ model. (left) Evolution of the on-site magnetization of the indicated sites using TIMES-ADAPT-I (circles), first-order Trotter (triangles), and exact evolution (solid lines). (right) The infidelity relative to the exact evolution for TIMES-ADAPT-I (circles), and first-order Trotter (triangles). The step size for the Trotter results was chosen to yield the same two-qubit gate depth as the unitary TIMES-ADAPT-I protocol. For clarity, only the even Trotter steps are shown in the plots.}
        \label{fig:WavePacket}
\end{figure*}

In Fig.~\ref{fig:WavePacket}, the real-time dynamics of the single-magnon wave packet in Eq.~\eqref{eq:GWavePacket} under $H_{\text{LFXXZ}}(-1.5,0.25)$ is shown. The plot on the left is the evolution of the site-wise magnetization $1-\langle\sigma_j^z\rangle$ of the indicated sites. The solid lines correspond to results from exact diagonalization, while the circles correspond to results using TIMES-ADAPT-I. The plot on the right shows the infidelity relative to the state obtained from exact diagonalization. We also present results using first-order Trotterization (triangles), where the step size $\delta t$ is chosen so that the two-qubit gate depth of the circuit matches that of TIMES-ADAPT-I, which is the sum of the depths for the state preparation (see Eq.~\eqref{eq:PrepState}) and the basis-change unitary $V_A(\vec{\theta}^*)$. An upper bound for the total two-qubit gate depth for all-to-all qubit connectivity comes out to $8m(N_s-1)+11N_{\text{adapt}}$, where $N_{\text{adapt}}$ is the number of layers of ADAPT-VQE. In our case, with $N_s=m=7$ and $N_{\text{adapt}}=7$, the two-qubit gate depth is $413$. We obtain this upper bound using two results: 
\begin{itemize}
    \item In Ref.~\cite{zindorf2025}, it was shown that any $n$-controlled SU(2) unitary can be implemented using a CNOT gate depth of $8n$ (for $n\geq6$). The state-preparation circuit in our case, as shown in Appendix~\ref{app:t0prep}, is a series of $m$ such gates, giving us a two-qubit depth of $8m(N_s-1)$ for the initial state preparation.
    \item We are using the QEB pool for the TETRIS-ADAPT-VQE protocol~\cite{AnsatasiouPRR2024} since it preserves $S_z$. In Ref.~\cite{YordanovNatComm2021}, it was shown that the more expensive double-excitation operators can be implemented using a circuit with a CNOT depth of $11$ in the case of all-to-all qubit connectivity. Assuming each of the $N_{\text{adapt}}$ layers of $V_A(\vec{\theta}^*)$ has at least one double-excitation operator, we get the upper bound of $11N_{\text{adapt}}$ for the two-qubit gate depth of the basis-change unitary for all-to-all connectivity.
\end{itemize}
The first-order Trotterization of $H_{\text{LFXXZ}}(J_z,h_z)$ is given by
\begin{equation}\label{eq:FOTrotter}
    e^{-i\,H_{\text{LFXXZ}}\,\delta t}\approx\left(e^{-i\,H_{\text{o}}\delta t}\,e^{-i\,H_{\text{e}}\delta t}e^{-i\,H_{\text{z}}\delta t}\right)^{N_t},
\end{equation}
where $H_{\text{o,e}}$ are, respectively, the interaction terms in the Hamiltonian in Eq.~\eqref{eq:HTFXXZ} starting on odd and even sites, while $H_{\text{z}}$ contains all the single-qubit terms. The total evolution time is $T=10.0$, with $N_t=68$ Trotter steps, corresponding to $\delta t\approx0.15$. The Trotter results diverge from the exact evolution over time, with a growing resource cost, whereas TIMES-ADAPT-I remains accurate for arbitrarily long times, while using a fixed-depth circuit. We do note that despite the worsening of the fidelity of the total state with time, the quality of the Trotter results are better for the local observable shown on the left panel of Fig.~\ref{fig:WavePacket} on the boundary site of the lattice than the same on the bulk site. This is due to the fact that the Trotter error accumulates at a higher rate on the bulk sites on account of it having two nearest neighbors, as opposed to just the one for the boundary sites. 

\subsection{Energy transport}
The problem of energy transport involves simulating the dynamics of a perturbed ground state of the system over long time scales. In this subsection, we will consider the evolution of a Gaussian perturbation on top of the ground state of the Heisenberg XXZ model with periodic boundary conditions, with a staggered external longitudinal magnetic field (SFXXZ) described by the Hamiltonian
\begin{multline}
     H_{\text{SFXXZ}}(J_z, h_z)=\sum_{k=1}^{N_s}\left(\sigma_{k}^x\sigma_{k+1}^x+\sigma_{k}^y\sigma_{k+1}^y\right.\\ \left.+J_z\sigma_{k}^z\sigma_{k+1}^z\right)+h_z\sum_{k=1}^{N_s}(-1)^k\sigma_k^z,
\end{multline}
where, $J_z$ is the coupling in the $Z$-direction, $h_z$ is the strength of the staggered magnetic field, $N_s$ is the number of spins in the chain, and $\sigma^{x,y,z}_j$ are the Pauli operators acting on qubit $j$. This model has been studied in the context of energy and thermal transport in~\cite{BulchandaniPNAS2023}. In this section, we study the dynamics under $H_{\text{SFXXZ}}(1.5,0.5)$. The initial perturbed state, which we will denote as $\ket{\psi_p(k_0,\sigma,0)}$ is the ground state of the perturbed Hamiltonian defined as
\begin{equation}
    H_g\equiv H_{\text{SFXXZ}}(1.5,0.5)+\delta H,
\end{equation}
where the perturbation $\delta H$ is $H_{\text{SFXXZ}}$ with inhomogeneous couplings given by a gaussian profile centered around the center of the lattice:
\begin{multline}
    \delta H=\sum_{k=1}^{N_s}J_k\left(\sigma_{k}^x\sigma_{k+1}^x+\sigma_{k}^y\sigma_{k+1}^y\right.\\ \left.+\sigma_{k}^z\sigma_{k+1}^z+(-1)^k\sigma_k^z\right),
\end{multline}
with
\begin{equation}
    J_k=\frac{1}{\sigma\sqrt{2\pi}}e^{-i\,(k-k_0)^2/(2\,\sigma^2)},
\end{equation}
where $k_0, \sigma$ are the mean and standard deviation of the distribution, respectively. The initial state $\ket{\psi_p(k_0,\sigma,0)}$ lives primarily (but not entirely) in the low-energy subspace of $H_{\text{SFXXZ}}$, making it a good test for TIMES-ADAPT-II. Since $H_{\text{SFXXZ}}$ and $H_g$ both commute with $S_z$, the evolution of the inhomogeneous initial state always remains in a subspace with a fixed value of $S_z=1.0$. We then use TIMES-ADAPT-II to study the dynamics of this state under $H_{\text{SFXXZ}}(J_z,h_z)$ using a fixed-depth circuit. 

\begin{figure*}
         \includegraphics[width=0.49\textwidth]{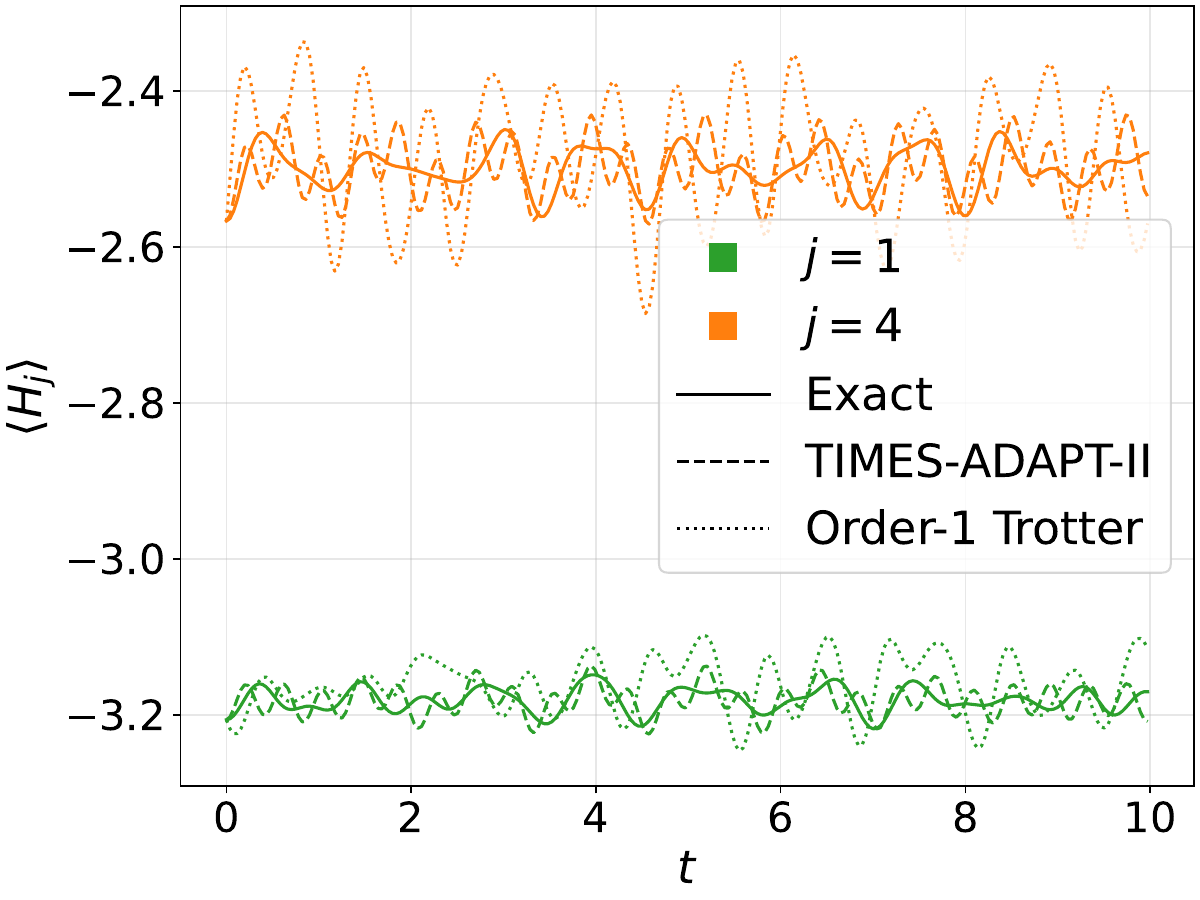}
         \label{fig:ETransport_Obs}
     \hfill
         \includegraphics[width=0.49\textwidth]{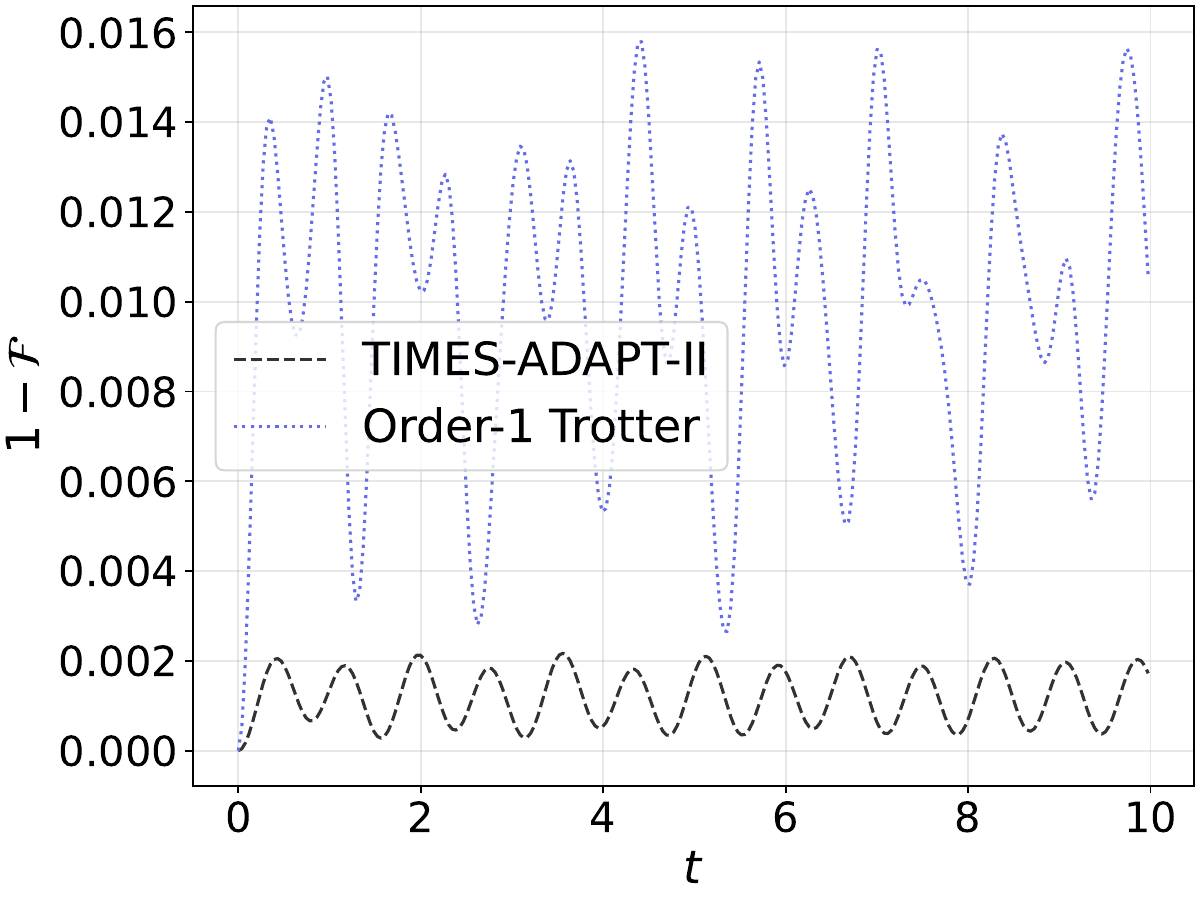}
         \label{fig:ETransport_InF}
        \caption{Time evolution of an inhomogeneous low-energy state in the staggered longitudinal-field XXZ model. (left) Evolution of the energy density on the indicated sites using TIMES-ADAPT-II (dashed lines), first-order Trotter (dotted lines), and exact evolution (solid lines). (right) The infidelity relative to the exact evolution for TIMES-ADAPT-II (dashed line), and first-order Trotter (dotted line). The step size for the Trotter results was chosen to yield the same two-qubit gate depth as the TIMES-ADAPT-II protocol.}
        \label{fig:ETransport}
\end{figure*}

In Fig.~\ref{fig:ETransport}, the energy transport of the perturbed state $\ket{\psi_p(4,1.0,0)}$ is shown under $H_{\text{SFXXZ}}(1.5,0.5)$. The plot on the left is the evolution of the on-site energy density
\begin{equation}
    H_k = \sigma^x_k\sigma^x_k+\sigma^y_k\sigma^y_k+J_z\sigma^z_k\sigma^z_k+(-1)^kh_z\sigma^z_k,
\end{equation}
where the site $N_s+1$ is the first site on the periodic lattice. The solid lines correspond to results from exact diagonalization, while the dashed lines correspond to results using TIMES-ADAPT-II. The plot on the right shows the infidelity relative to the result from exact diagonalization. We also present results using first-order Trotterization (dotted lines), where the step size $\delta t$ is chosen so that the two-qubit gate depth of the circuit (all-to-all connectivity) matches that of the unitary $\tilde{U}(t)$ in Eq.~\eqref{eq:TIMES-ADAPT-II-Unitary} applied in TIMES-ADAPT-II. Note that there is no state-preparation cost here. The total two-qubit gate depth for all-to-all connectivity is $8m(N_s-1)+2\times11N_{\text{adapt}}$, where $N_{\text{adapt}}$ is the number of layers of ADAPT-VQE. The first term in the depth is the cost of the diagonal unitary $D(t)$ in Eq.~\eqref{eq:DtTIMES-II}, which can be broken down into a series of $m$ multi-controlled $SU(2)$ unitaries. The second term in the cost comes from conjugating $D(t)$ with the basis-change unitary $V_A(\vec{\theta}^*)$. In our case, with $N_s=7$ and $m=4$, the total two-qubit gate depth is $1534$. While in terms of gates, TIMES-ADAPT-II is more expensive, we gain the advantage of not having to know the coefficients of the initial state in the truncated eigenbasis. The Trotterization is done the same way as indicated in Eq.~\eqref{eq:FOTrotter}, with $H_z$ being the staggered magnetic field in this case. The total evolution time is $T=10.0$, with $N_t=255$.

From the right panel of Fig.~\ref{fig:ETransport}, we see that the infidelity of the evolution generated by TIMES-ADAPT-II remains low out to large times ($1-\mathcal{F}\sim\mathcal{O}(10^{-3})$), whereas that of first-order Trotterization grows as much as an order of magnitude larger. A similar conclusion can be arrived at using the observables shown in the left panel of Fig.~\ref{fig:ETransport}. While the evolution of the observable obtained from TIMES-ADAPT-II does not exactly match the exact diagonalization results, it remains in its vicinity. This could be further improved by using a larger cutoff for the number of eigenstates found by TEPID-ADAPT. The Trotter results on the other hand show a much larger variance in the value of the observable.

\section{Initial state}\label{app:t0prep}
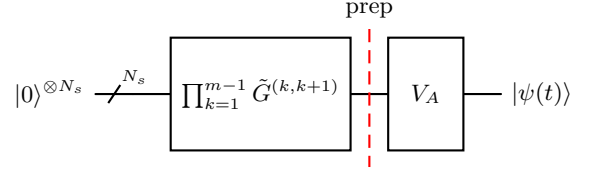
\begin{figure}[ht!]
    \centering
    \begin{quantikz}[classical gap=0.07cm]
        \lstick{$\ket{0}^{\otimes N_s}$} & \qwbundle{N_s} & \gate[][1cm][1.5cm]{\prod_{k=1}^{m-1}\tilde{G}^{(k,k+1)}}\slice{prep}&\gate[][1cm][1.5cm]{V_A}&\rstick{$\ket{\psi(t)}$}
    \end{quantikz}
    \caption{Block diagram for the preparation of the time-evolved state. The state after the dotted line is $\ket{\chi(t)}$ in Eq.~\eqref{eq:PrepState}. Following this, there is an action of $V_A$ to prepare the time-evolved state. Each modified Givens rotation has parameters $(\theta_k(t),\varphi_k)$.}
    \label{fig:PrepCircuit}
\end{figure}

In this Appendix, we prescribe a way to prepare the state $\ket{\chi(t)}$, defined in Eq.~\eqref{eq:PrepState} and restated below:
\begin{equation*}
	\ket{\chi(t)}=\sum_{k=1}^m\alpha_k\,e^{-i\,E_k\,t}\ket{c_k}.
\end{equation*}
We will again assume that it lives in the subspace spanned by the truncated eigenbasis and that we know its corresponding coefficients $\{\alpha_k\}_{k=1}^m$. We start with the state $\ket{0}^{\otimes N_s}$ and apply a series of modified Givens rotations
\begin{multline}
	    \tilde{G}^{(k,k+1)}(\theta_k,\varphi_k)\equiv \exp\left(i\,\theta_{k+1}\,P^{(k)}\right)\times\\\exp\left({i\,\varphi_k\gamma^{(k,k+1)}}\right)\exp\left(-i\,\theta_{k+1}\,P^{(k)}\right),
\end{multline}
where
\begin{equation}
	\left[\gamma^{(k,k+1)}\right]_{x,y}=i\left(\delta_{x,c_k}\,\delta_{y,c_{k+1}}-\delta_{x,c_{k+1}}\,\delta_{y,c_{k}}\right),
\end{equation}
and
\begin{equation}
    \left[P^{(k)}\right]_{x,y}=\delta_{x,c_k}\,\delta_{y,c_k},
\end{equation}
using $\left[\,\cdot\,\right]_{x,y}$ to denote the $(x,y)^{th}$ matrix element and $\delta_{i,j}$ for a Kronecker delta. $\gamma$ generates a rotation between two computational basis elements, while $P$ generates a phase on specific basis elements. The subscript $c_k$ denotes a computational basis element. For example, $\ket{1101}$ would correspond to the subscript $13$ in the Kronecker deltas. Geometrically this series of modified Givens rotations corresponds to a rotation from $\left(1,0,\ldots,0\right)$ to a general complex unit vector
\begin{equation}
    \left(1,0,\ldots,0\right)\to\left(\alpha_1\,e^{-i\,E_1\,t},\ldots,\alpha_m\,e^{-i\,E_m\,t},0,\ldots,0\right),
\end{equation}
up to an overall phase. Here, the vectors are in the computational basis, and for sake of presentation, we have reordered the computational basis so that the coefficients of $\{\ket{c_k}\}_{k=1}^m$ appear in the first $m$ entries of the vector above. The remaining components correspond to a padding with $2^{\lceil\log_2m\rceil}-m$ zeros. 
Next, we write down the gate angle parametrization in terms of the coefficients $\{\alpha_k\}_{k=1}^m$. 
First, let us transform the complex coefficients into their polar form, and rewrite Eq.~\eqref{eq:PrepState}:
\begin{align}
    \alpha_k & \equiv r_k\,e^{i\,\zeta_k},\nonumber\\
    \ket{\chi(t)}&=\sum_{k=1}^m{r_k}e^{i(\zeta_k-E_k\,t)}\ket{c_k}.
\end{align}
Introducing the notation $\Delta[\cdot]_k\equiv[\cdot]_k-[\cdot]_1$, we factor out and discard a global phase $\zeta_1-E_1t$ to obtain
\begin{equation}
    \ket{\chi(t)}\simeq\sum_{k=1}^m{r_k}e^{i(\Delta\zeta_k-\Delta E_k\,t)}\ket{c_k}.
\end{equation}
The energy differences $\Delta E_k\equiv (E_k-E_1)$ are readily obtained from the converged parameters of TEPID-ADAPT.
The phase parameters $\theta_k$ entering into our modified Givens rotations are thus
\begin{equation}
    \theta_k(t) \equiv 
    \begin{cases}
        0 & k=1\\\\
        \xi_{k-1}(t)-\xi_k(t) & k>1
    \end{cases},
\end{equation}
where $\xi_k(t)=\Delta\zeta_k-\Delta E_k\,t$. These can be interpreted as the phases of each coefficient $\alpha_k$ (factoring out a global phase), dressed by a time-dependent term accounting for the precession of each eigenvector.
Meanwhile, the rotation parameters $\varphi_k$ are related to the amplitudes $r_k\equiv|\alpha_k|$
by the hyperspherical coordinate transformations
\begin{equation}\label{eq:PolarParam-mag}
    \sqrt{r_j}=
    \begin{cases}
        \left(\prod_{k=1}^{j-1}\sin(\varphi_k)\right)\cos(\varphi_j) & 1\leq j<m \\\\
        \left(\prod_{k=1}^{m-1}\sin(\varphi_k)\right) & j=m
    \end{cases}.
\end{equation}
If we know the coefficients of $\ket{\psi(0)}$ in the eigenbasis $\{\alpha_k\}_{k=1}^m$, we can invert the transformation in Eq.~\eqref{eq:PolarParam-mag}, and apply the circuit in Fig.~\ref{fig:PrepCircuit} to obtain the time-evolved state at an arbitrary time.

\subsection{Determining the coefficients}\label{app:FindCoeffs}
So far, we have assumed knowledge of the coefficients of the initial state $\ket{\psi(0)}$ in the truncated eigenbasis. In this subsection, we provide a method to determine them, assuming we have a circuit to prepare $\ket{\psi(0)}$. We run a variational quantum algorithm that uses the circuit in Fig.~\ref{fig:PrepCircuit} with the angles $(\vec{\theta}(0),\vec{\varphi})$ as variational parameters. The cost function for this would be the infidelity between the variationally prepared state $\ket{\phi(\vec{\theta}(0),\vec{\varphi})}$ and $\ket{\psi(0)}$
\begin{equation}
    C(\vec{\theta}(0),\vec{\varphi})=1-\lvert\bra{\psi(0)}\phi(\vec{\theta}(0),\vec{\varphi})\rangle\rvert^2.
\end{equation}
This can be measured efficiently using a swap test~\cite{BarencoSIAMJC1997,BuhrmanPRL2001} or ``compute-uncompute'' methods~\cite{HavlicekNature2019}. If the cost function vanishes upon convergence, we know that $\ket{\psi(0)}$ is in the truncated eigenspace. If not, we can estimate the error in our time-evolution, as we show in Appendix~\ref{app:ErrorEstimates}. 


\section{Error Analysis}\label{app:ErrorEstimates}
In this Appendix, we write down the errors as a function of time when the initial state $\ket{\psi(0)}$ is not entirely in the subspace found by TEPID-ADAPT for both our methods. Below, we consider an initial state $\ket{\psi(0)}$ in the subspace $\mathbf{S}'$ spanned by eigenstates $\left\{\ket{\psi_k}\right\}_{k\in\mathbf{S}'}$ of a Hamiltonian, $H$:
\begin{equation}
    \ket{\psi(0)}=\sum_{k\in\mathbf{S}'}\alpha_k\ket{\psi_k}.
\end{equation}
Let $\mathbf{S}\subseteq\mathbf{S}'$ be the subspace found by TEPID-ADAPT spanned by the eigenstates $\left\{\ket{\psi_k}\right\}_{k\in\mathbf{S}'}$.


\subsection{TIMES-ADAPT-I}
With TIMES-ADAPT-I, we effectively evolve the projection of $\ket{\psi(0)}$ onto $\mathbf{S}$
\begin{equation}
    \ket{\psi_{\mathbf{S}}(0)}\sim\mathbb{P}_{\mathbf{S}}\ket{\psi(0)}=\sum_{k\in\mathbf{S}}\alpha_k\ket{\psi_k},
\end{equation}
where $\mathbb{P}_{\mathbf{S}}$ is the projector onto the subspace $\mathbf{S}$. The corresponding time-evolved states are
\begin{align}
    \ket{\psi(t)}&=\sum_{k\in\mathbf{S'}}\alpha_k\,e^{-i\,E_k\,t}\ket{\psi_k},\\
    \ket{\psi_{\mathbf{S}}(t)}&=\sum_{k\in\mathbf{S}}\alpha_k\,e^{-i\,E_k\,t}\ket{\psi_k}.
\end{align}
The fidelity between these two states is
\begin{multline}
    \mathcal{F}(t)\equiv\lvert\bra{\psi_{\mathbf{S}}(t)}\psi(t)\rangle\rvert^2\\=\Bigg\lvert\sum_{l\in\mathbf{S'},\,k\in\mathbf{S}}\alpha^*_k\,\alpha_l\,e^{-i\,(E_l-E_k)\,t}\langle\psi_k\ket{\psi_l}\Bigg\rvert^2,
\end{multline}
where the wave overlap will yield a Kronecker delta due to orthogonality of the eigenstates
\begin{equation}
    \mathcal{F}(t)=\left(\sum_{k\in\mathbf{S}}\lvert\alpha_k\rvert^2\right)^2.
\end{equation}
The fidelity has no dependence on time, and therefore knowing the initial state fidelity is enough to know the fidelity at all times. This quantity is the minimum of the cost function in the variational quantum algorithm used to find the coefficients in the eigenbasis, described in Appendix~\ref{app:FindCoeffs}.

The time-evolved state in the subspace $\mathbf{S}$ found by TEPID-ADAPT is
\begin{equation}
    \ket{\psi_{\mathbf{S}}(t)}=\sum_{k\in\mathbf{S}}\alpha_k\,e^{-i\,\tilde{E}_k\,t}\ket{\psi_k},
\end{equation}
where the effective energies are of the effective Hamiltonian $\tilde{H}$ introduced in Sec.~\ref{sec:OurMethod}. These are given explicitly in Eq.~\eqref{eq:EffEnergies}.

\subsection{TIMES-ADAPT-II}
With TIMES-ADAPT-II, we evolve the full initial state, $\ket{\psi(0)}$, by the effective Hamiltonian, $\tilde{H}$, introduced in Sec.~\ref{sec:OurMethod} with energies
\begin{equation}\label{eq:EffEnergies}
    \tilde{E}_k=
    \begin{cases}
        E_k & k\in\mathbf{S} \\\\
        0   & k\in\mathbf{S}'\setminus\mathbf{S}
    \end{cases},
\end{equation}
giving us the state
\begin{equation}
    \ket{\psi_{\mathbf{S}}(t)}=\sum_{k\in\mathbf{S}}\alpha_k\,e^{-i\,E_k\,t}\ket{\psi_k}+\sum_{k\in\mathbf{S}'\setminus\mathbf{S}}\alpha_k\,\ket{\psi_k}.
\end{equation}
Here, for the sake of analysis, we are choosing to expand the portion of the state in $\mathbf{S'}\setminus\mathbf{S}$ in the eigenbasis of $H$. These eigenstates do not evolve with time. The fidelity with the state evolved under the full Hamiltonian is
\begin{multline}
    \mathcal{F}(t)\equiv\lvert\bra{\psi_{\mathbf{S}}(t)}\psi(t)\rangle\rvert^2\\=\Bigg\lvert\sum_{l,k\in\mathbf{S}'}\alpha^*_k\,\alpha_l\,e^{-i\,(E_l-\tilde{E}_k)\,t}\langle\psi_k\ket{\psi_l}\Bigg\rvert^2,
\end{multline}
where the wave overlap will yield a Kronecker delta due to the orthogonality of the eigenstates. Using Eq.~\eqref{eq:EffEnergies} the fidelity simplifies to:
\begin{equation}
    \mathcal{F}(t)=\Bigg\lvert\sum_{k\in\mathbf{S}}\lvert\alpha_k\rvert^2+\sum_{k\in\mathbf{S'}\setminus\mathbf{S}}\lvert\alpha_k\rvert^2\,e^{-i\,E_k\,t}\Bigg\rvert^2.
\end{equation}
Unlike TEPID-ADAPT-I, the infidelity now depends on \textit{all} components of the initial state $\ket{\psi(0)}$ (not just those in $\mathbf{S}$). The fidelity has an oscillatory behavior with the frequencies set by the energies of the eigenstates not in $\mathbf{S}$. As a result, this fidelity can either exceed or fall below that of TIMES-ADAPT-I depending on the time we are interested in.


\end{document}